\documentclass[prb,showpacs,preprintnumbers,amsmath,superscriptaddress,twocolumn,citeautoscript]{revtex4} %
\usepackage{amsfonts}
\usepackage{mathrsfs}

\usepackage{graphicx}
\usepackage{epstopdf}
\usepackage{dcolumn}
\usepackage{bm}
\usepackage{amsmath}

\makeatletter
\renewcommand{\section}{\@startsection{section}{1}{0mm}
  {-\baselineskip}{0.5\baselineskip}{\bf\leftline}}
\makeatother

\makeatletter
\renewcommand{\subsection}{\@startsection{subsection}{1}{0mm}
  {-\baselineskip}{0.5\baselineskip}{\bf\leftline}}
\makeatother

\makeatletter
\renewcommand{\subsubsection}{\@startsection{subsubsection}{1}{0mm}
  {-\baselineskip}{0.5\baselineskip}{\bf\;\leftline}}%
\makeatother

\begin{document}

\preprint{APS/123-QED}
\title{Tight binding analysis of Si and GaAs ultra thin bodies with subatomic resolution }

\author{Yaohua P.~Tan}\email{tyhua02@gmail.com}
\affiliation{%
School of Electrical and Computer Engineering,Network for
Computational Nanotechnology,Purdue University, West Lafayette,
Indiana,USA,47906 
}

\author{Michael~Povolotskyi}
\affiliation{%
School of Electrical and Computer Engineering,Network for
Computational Nanotechnology,Purdue University, West Lafayette,
Indiana,USA,47906 
}
\author{Tillmann~Kubis }
\affiliation{%
School of Electrical and Computer Engineering,Network for
Computational Nanotechnology,Purdue University, West Lafayette,
Indiana,USA,47906 
}
\author{Timothy B.~Boykin} \affiliation{
Department of Electrical and Computer Engineering, University of Alabama in Huntsville,Huntsville, Alabama 35899 USA 
}%
\author{Gerhard~Klimeck }
\affiliation{%
School of Electrical and Computer Engineering,Network for
Computational Nanotechnology,Purdue University, West Lafayette,
Indiana,USA,47906 
}

\date{\today}
\begin{abstract}
Empirical tight binding(ETB) methods are widely used in atomistic
device simulations. Traditional ways of generating the ETB
parameters rely on direct fitting to bulk experiments or theoretical
electronic bands. However, ETB calculations based on existing
parameters lead to unphysical results in ultra small structures like
the As terminated GaAs ultra thin bodies(UTBs). In this work, it is
shown that more reliable parameterizations can be obtained by a
process of mapping \textit{ab-initio} bands and wave functions to
tight binding models. This process enables the calibration of not
only the ETB energy bands but also the ETB wave functions with
corresponding \textit{ab-initio} calculations. Based on the mapping
process, ETB model of Si and GaAs are parameterized with respect to
hybrid functional calculations. Highly localized ETB basis functions
are obtained. Both the ETB energy bands and wave functions with
subatomic resolution of UTBs show good agreement with the
corresponding hybrid functional calculations. The ETB methods can
then be used to explain realistically extended devices in
non-equilibrium that can not be tackled with \textit{ab-initio}
methods.
\end{abstract}

\maketitle


\section{Introduction}

Modern semiconductor nanodevices have reached critical device
dimensions in the sub-10 nanometer range. These devices consist of
complicated two or three dimensional geometries and are composed of
multiple materials. Confined geometries such as ultra thin body
(UTB)\cite{YKChoi_UTBFETs}, FinFETs\cite{Hismoto_FinFET} and
nanowires\cite{Xuan_Nanowire} structures are usually adopted in
nanometer scale device designs to obtain desired performance
characteristics.Most of the electrically conducting devices are not
arranged in infinite periodic arrays, but are of finite extent with
contacts controlling the current injections and potential
modulation.  Typically, there are about 10000 to 10 million atoms in
the active device region with contacts controlling the current
injection. These finite sized structures suggest an atomistic, local
and orbital-based electronic structure representation for device
level simulation. Quantitative device design requires the reliable
prediction of the materials' band gaps and band offsets within a few
meV and important effective masses within a few percent in the
geometrically confined active device regions. ETB model is usually
fitted to bulk dispersions without any definition of the spatial
wave function details. However, recent \textit{ab-initio} study of
UTBs~\cite{Ryan_Hatcher_UTBs} showed that the surface carrier
distribution in confined systems is strongly geometry and material
dependent. This suggests that the charge distribution for realistic
predictions of nanodevice performances should be resolved with
subatomic resolution.

\textit{Ab-initio} methods offer atomistic representations with
subatomic resolution for a variety of materials. However, accurate
\textit{ab-initio} methods, such as Hybrid functionals~\cite{HSE06},
GW~\cite{Hybertsen_GW} and BSE approximations~\cite{BSE_PRL} are in
general computationally too expensive to be applied to systems
containing millions of atoms. Furthermore, those methods assume
equilibrium  and cannot truly model out-of-equilibrium device
conditions where e.g. a large voltage might have been applied to
drive carriers. The ETB methods are numerically much more efficient
than \textit{Ab-initio} methods. ETB has established itself as the
standard state-of-the-art basis for realistic device simulations.
It has been successfully applied to electronic structures of
millions of atoms~\cite{Klimeck_QuantumDot} as well as on
non-equilibrium transport problems that even involve inelastic scattering.~\cite{Klimeck_RTD}%
The accuracy of the ETB methods depend critically on the careful
calibration of the empirical parameters. The traditional way to
determine the ETB parameters is to fit ETB band structures to
experimental data of bulk
materials.~\cite{Jancu_Tightbinding,Boykin_TB_strain}

\begin{figure}[h] \center
\includegraphics[width = \columnwidth]{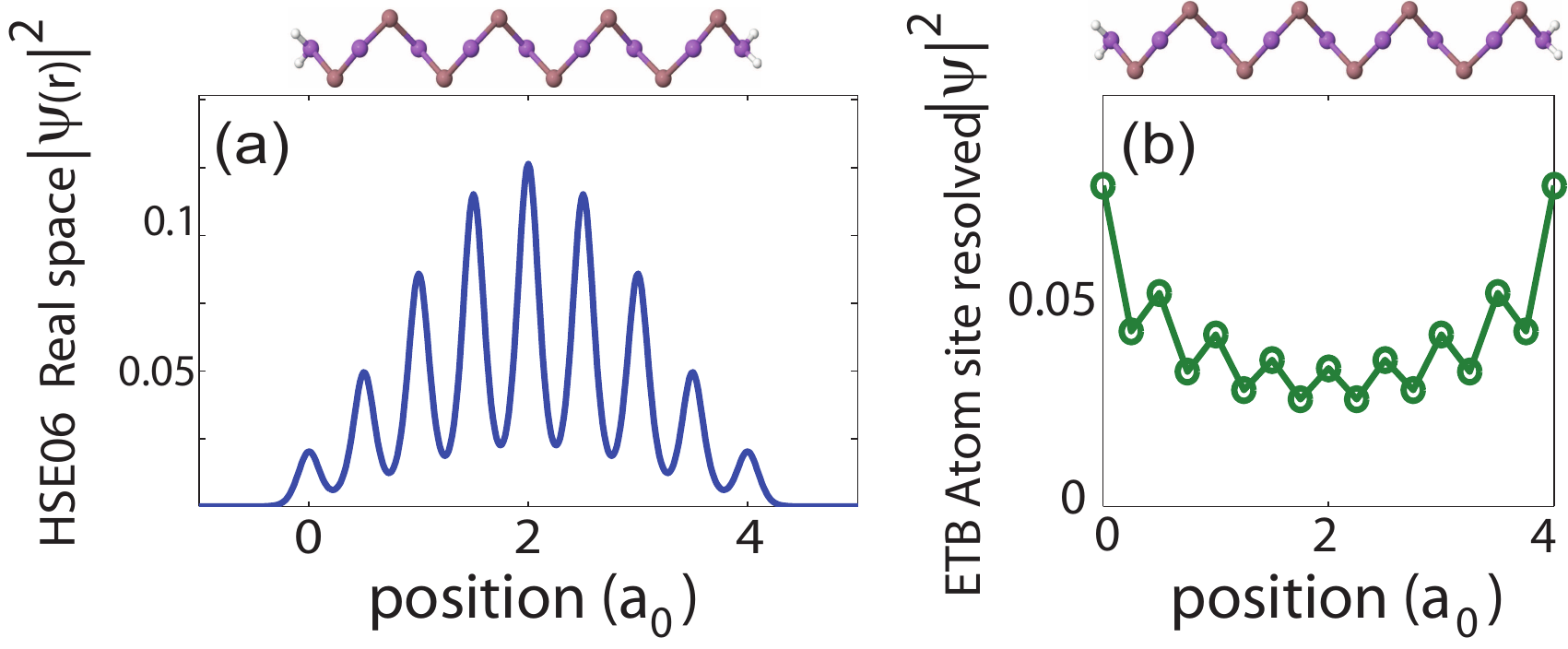}
.\caption{In As terminated GaAs UTBs, hybrid functional probability
amplitudes (a) of the top valence bands are confined states with
probability amplitude
peaking in the center of the UTB, while the ETB valence states (b) are surface states. }%
\label{fig:unphysical_TB_states}%
\end{figure}

The ETB basis functions remain implicitly defined during traditional
fitting processes. The lack of explicit basis functions makes it
difficult to predict wave function dependent quantities like optical
matrix elements with high precision. More importantly, ETB models
parameterized by traditional fitting processes suffer from potential
ambiguity when applied to ultra small structures such as UTBs,
nanowires and more complicated geometries. For instance, the
existing ETB parameters of GaAs~\cite{Boykin_TB_strain} applied to a
As terminated GaAs UTB with an implicit Hydrogen passivation
model~\cite{TB_passivation} results in unphysical top valence band
states as shown in Fig.~\ref{fig:unphysical_TB_states}: The real
space probability amplitudes of \textit{ab-initio} topmost valence
bands correspond to confined states with the probability amplitude
peaking in the center of the UTB rather than the surface of the UTB as in ETB.
In Fig.~\ref{fig:unphysical_TB_states}, the
hybrid functional calculations include Hydrogen atoms explicitly
whereas the ETB calculations include only their impact
implicitly.~\cite{TB_passivation} The mismatch between the envelopes
of ETB and \textit{ab-initio} wavefunctions suggests a calibration
of wave functions in the ETB parameterization process is necessary.
It is also found that the method of passivation (i.e. implicit or
explicit inclusion of Hydrogen atoms) has an effect on the nature of
the valence band states.

Therefore, a more fundamental fitting process that relates both the
band structure and the wave functions of ETB models with
\textit{ab-initio} calculations is desirable. Existing approaches to
construct localized basis functions and tightbinding-like Hamiltonians from
\textit{ab-initio} results include maximally localized Wannier
functions(MLWF),~\cite{Marzari_Wannier_functions,
MLWF_entangled_bands_Souza} quasi-atomic
orbitals,~\cite{Qian_quasiatomic_orbitals,Lu_quasiatomic_orbitals}
or DFT-TB analysis.~\cite{Urban_TBfromDFT}The MLWFs are constructed
using Bloch states of either isolated
bands~\cite{Marzari_Wannier_functions} or entangled
bands.~\cite{MLWF_entangled_bands_Souza}These methods typically
include interatomic interactions beyond first nearest neighbors.
However, these methods do not eliminate the above discussed
ambiguity of the commonly used orthogonal sp3d5s* ETB models.
Furthermore, these approaches usually disregard excited orbitals
which are often needed to correctly parameterize semiconductors
conduction bands. In previous work, it was already suggested how to
generate ETB parameters that are compatible with typical ETB models
and still reproduce \textit{ab-initio}
results.~\cite{DFTMapping_JCEL}This previous method was already
applied to several materials such as GaAs,
MgO~\cite{DFTMapping_JCEL} and SmSe~\cite{SmSe_Zhengping} and
yielded a good agreement between bulk ETB and \textit{ab-initio} band
structures. However, the resulting wave functions did not
satisfactorily agree with the \textit{ab-initio} wave functions.

In this paper, an improved algorithm of
Ref.~\onlinecite{DFTMapping_JCEL} is presented that "maps"
\textit{ab-initio} results (i.e. eigenenergies and eigenfunctions)
to tight binding models. Compared with the previous
work\cite{DFTMapping_JCEL}, the presented method allows much better
agreement of the ETB and \textit{ab-initio} wave functions.
In this present mapping algorithm, rigorous, wavefunction-derived ETB
parameters for the Hamiltonian, for highly localized basis
functions, and for explicit surface passivation are obtained. It is
important to mention that the ETB Hamiltonian of this method can be
limited to first nearest neighbor interaction. The mapping process
is applied to both bulk Si and GaAs to generate ETB parameters and
explicit basis functions from corresponding hybrid functional
calculations. It is demonstrated in this work, that the
wave-function derived ETB Hamiltonian does not yield the ambiguity
discussed with Fig.~\ref{fig:unphysical_TB_states}. In the same way,
the transferability of the ETB model to nanostructures is improved.
This is demonstrated by a comparison of ETB and Hybrid functional
results in GaAs and Si UTBs.

This paper is organized as follows. In section~\ref{sec:method}, the
algorithm of parameter mapping from \textit{ab-initio} calculations
to tight binding models is described. Section~\ref{sec:results}
shows the application of the mapping algorithm to bulk and UTB
systems. Subsection~\ref{sec:application_to_bulk_materials} presents
the application of the present algorithm to bulk Si and GaAs. Bulk
band structures and realspace basis functions are shown and
discussed there as well. Subsection~\ref{sec:application_to_UTB}
shows the application of the algorithm to UTB systems and compares
ETB band structures and wave functions with corresponding
\textit{ab-initio} results. The algorithm and its results are
summarized in Section~\ref{Sec:conclusion}.

\section{Method\label{sec:method}}

\subsection{Parameter Mapping Algorithm}
\begin{figure}[ptb] \center
\includegraphics[width = 8cm]{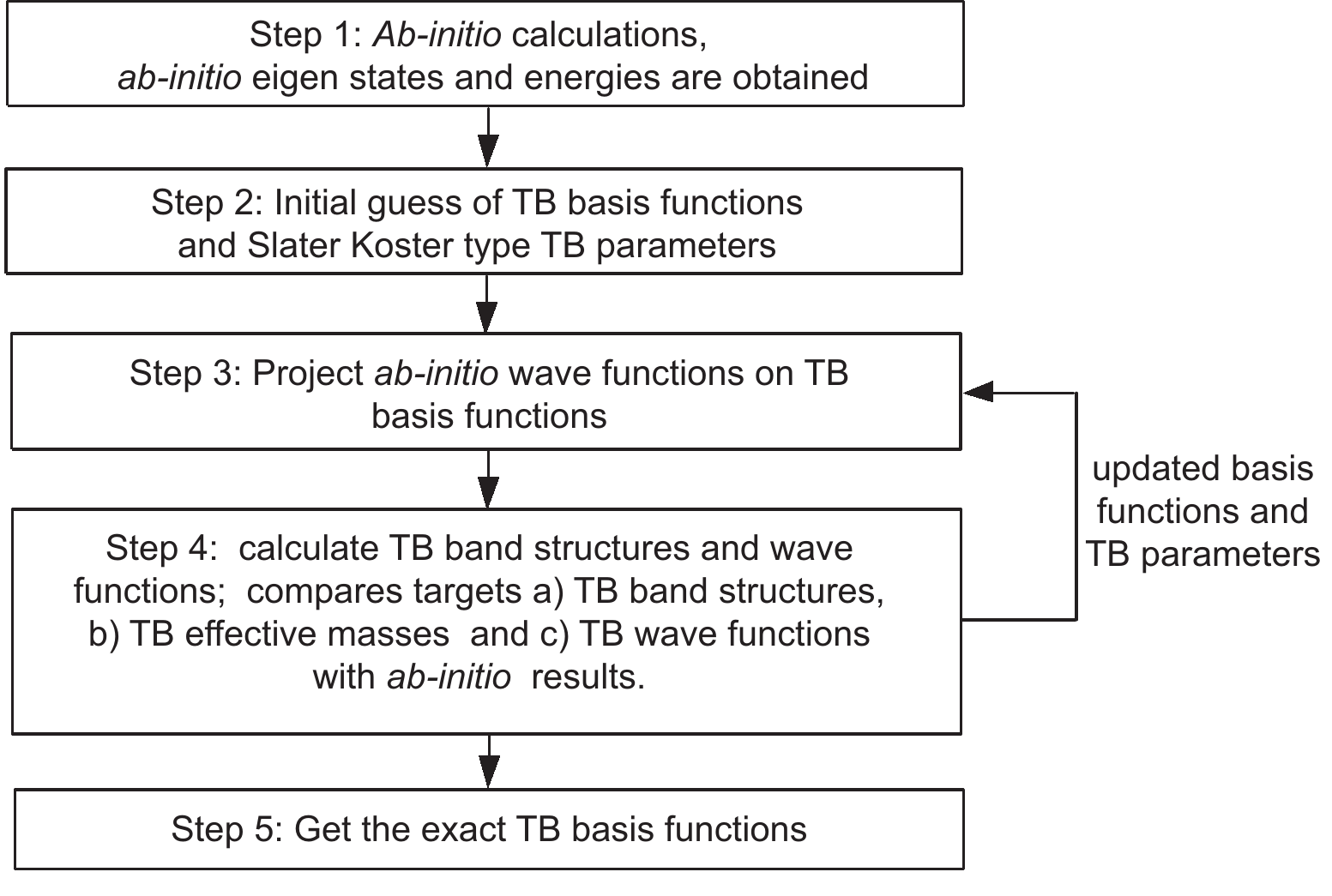}
.\caption{The process of mapping from \textit{ab-initio}
calculations to Tight Binding by which the ETB parameters and ETB
basis functions are extracted iteratively. }%
\label{fig:Process_flow}%
\end{figure}

The algorithm of the parameter mapping from \textit{ab-initio}
results to ETB models is shown in Fig.~\ref{fig:Process_flow}.
As will be shown in the following, the ETB parameters and basis
functions are obtained in an iterative fitting procedure that spans
over 5 steps (with steps 3 through 4 being iterated). The resulting
1st nearest neighbor Hamiltonian $\hat{H}^{TB}\left(
\mathbf{k}\right)$ is of Slater Koster table
type.~\cite{Slater_Tightbinding,Podolskiy_TBElements} The resulting
basis $\mathfrak{B}_{\text{final}}$ is composed of orthonormal real
space functions $\mathfrak{B}_{\text{final}}=\left\{  \Psi_{n,l,m}%
^{\text{final}}\left(  \mathbf{r}\right)  \right\}  $ which have the
shape
(vectors are given in bold type)%
\begin{eqnarray}
  \nonumber \Psi_{a,n,l,m}\left(  \mathbf{r}\right) &=& \bar{Y}_{l,m}\left( \theta
,\phi\right)  \bar{R}_{a,n,l}\left(  r\right) + \\
   & & \sum_{\substack{l^{\prime },m^{\prime}\\\left(
l^{\prime},m^{\prime}\right)  \neq\left(  l,m\right)
}}\bar{Y}_{l^{\prime},m^{\prime}}\left(  \theta,\phi\right)  \tilde
{R}_{a,n,l,l^{\prime},m^{\prime}}\left(  r\right) .\quad
\label{eq:definition_atomic_orbitals_final}
\end{eqnarray}
Here, $a$ labels the atom type, whereas the $n$, $l$ and $m$ are
principle, angular and magnetic quantum numbers, respectively. All
materials considered in this work contain no magnetic polarization.
Therefore, the basis functions are spin independent. The tesseral
spherical harmonics $\bar{Y}_{l,m}\left( \theta,\phi\right)  $
describe the dependence of the basis functions on the angular
coordinates $\theta$ and $\phi$. The functions
$\bar{R}_{a,n,l}\left( r\right) $ and
$\tilde{R}_{a,n,l,l^{\prime},m^{\prime}}\left(  r\right)  $ define
the radial $r$ dependence of the basis functions. The contribution
of $\tilde{R}_{a,n,l,l^{\prime},m^{\prime}}$ to the basis functions
is much smaller than the contribution of $\bar{R}_{a,n,l}$. The
detailed shapes of the radial functions $\bar{R}_{a,n,l}\left(
r\right)  $ and $\tilde {R}_{a,n,l,l^{\prime},m^{\prime}}\left(
r\right) $ are subject to the fitting algorithm.

\textbf{Step 1:} First, electronic band structures $\varepsilon_{j}%
^{Ab}\left(  \mathbf{k}\right)  $ and wave functions $\psi_{j,\mathbf{k}}%
^{Ab}$ are solved which serve as fitting targets to the overall
mapping algorithm%
\begin{equation}
\hat{H}^{Ab}\left(  \mathbf{k}\right)  \left\vert \psi_{j,\mathbf{k}}%
^{Ab}\right\rangle =\varepsilon_{j}^{Ab}\left(  \mathbf{k}\right)
\left\vert \psi_{j,\mathbf{k}}^{Ab}\right\rangle .
\end{equation}
The index $j$ corresponds to the band index and $\mathbf{k}$
represents a momentum vector in the first Brillouin zone. In
principle, any method that is capable of solving band diagrams and
explicit basis functions can provide these fitting targets.
Throughout this work, however, hybrid functional calculations are
performed for step 1.~\cite{Vasp_HSE}

\textbf{Step 2:} In the second step, initial guesses for the ETB
basis functions and ETB parameters are defined. During the fitting
process, the ETB basis $\mathfrak{B}_{\text{initial}}$ is spanned by
non-orthogonal functions $\left\{  \Phi_{a,n,l,m}\left(
\mathbf{r}\right)  \right\}  $ given by%
\begin{equation}
\Phi_{a,n,l,m}\left(  \mathbf{r}\right) = \bar{Y}_{l,m}\left( \theta
,\phi\right)  R_{a,n,l}\left(  r\right).
\label{eq:definition_atomic_orbitals}%
\end{equation}
The $R_{a,n,l}\left(  r\right)$ in
Eq.~(\ref{eq:definition_atomic_orbitals}) differ from the
$\bar{R}_{a,n,l}\left(  r\right) $ of the final basis functions in
Eq.~(\ref{eq:definition_atomic_orbitals_final}). 
The details of the initial guesses for the diagonal and off-diagonal
elements of the Hamiltonian $\hat{H}^{TB}\left( \mathbf{k}\right)  $
are not essential for the overall algorithm. Nevertheless, initial
guesses that follow the framework of existing ETB parameter sets
improve the overall fitting convergence. Urban et al. and Lu et al.
discuss that interactions up to third nearest neighbors might be
needed to exactly reproduce \textit{ab-initio}
results.~\cite{Urban_TBfromDFT,Lu_quasiatomic_orbitals}In contrast,
we find that the interatomic interaction elements of
$\hat{H}^{TB}\left( \mathbf{k}\right)  $ can be limited to first
nearest neighbor interactions throughout this work while still
reproducing \textit{ab-initio} results very well. 

\textbf{Step 3:} The nonorthogonal basis functions
$\Phi_{a,n,l,m}\left( \mathbf{r}\right)  $ in position space are
transformed into the Bloch
representation~\cite{Ashcroft}$\Phi_{a,n,l,m,\mathbf{k}}\left(  \mathbf{r}%
\right)  $%
\begin{eqnarray}
  \nonumber \left\vert \Phi_{\alpha,\mathbf{k}}\right\rangle & \equiv & \Phi
_{a,n,l,m,\mathbf{k}}\left(  \mathbf{r}\right)   \\
   & = & \sum_{\mathbf{R}}\exp\left[ {i\mathbf{k}\cdot}\left(
{\mathbf{R}+\boldsymbol{\tau}}_{a}\right)  \right]
\Phi_{a,n,l,m}\left(
\mathbf{r}-\mathbf{R}-\boldsymbol{\tau}_{a}\right)  ,
\label{eq:bloch_basis_Tight_Binding}
\end{eqnarray}
where $\mathbf{\tau}_{a}$ is the position of atom type $a$ in the
unit cell and the sum runs over all unit cells of the system with
$\mathbf{R}$, the position of the respective cell. To improve
readability of all formulas in the Dirac notation, the indices of
atom type and quantum numbers are merged into Greek indices
$\alpha=\left(  a,n,l,m\right)  $. For the further steps, an
orthogonal basis $\mathfrak{B}_{\text{ortho}}=\left\{  \left\vert
\Psi _{\alpha,\mathbf{k}}\right\rangle \right\}  $ is created out of
the basis $\mathfrak{B}_{\text{initial}}$ with L\"{o}wdin's
symmetrical orthogonalization algorithm.~\cite{Lowdin_orth}Since
steps 4 and 5 are formulated in the basis
$\mathfrak{B}_{\text{ortho}}$, the wave functions $\left\vert
\psi_{j,\mathbf{k}}^{Ab}\right\rangle $ of step 1 must be
transformed into this basis%
\begin{equation}
\left\vert \psi_{j,\mathbf{k}}^{Ab}\right\rangle
\approx\hat{P}\left( \mathbf{k}\right)  \left\vert
\psi_{j,\mathbf{k}}^{Ab}\right\rangle
=\sum_{\alpha}c_{j,\alpha}\left(  \mathbf{k}\right)  \left\vert \Psi
_{\alpha,\mathbf{k}}\right\rangle , \label{eq:approx_abinit}%
\end{equation}
where
\begin{equation}
c_{j,\alpha}\left(  \mathbf{k}\right)  =\left\langle \Psi_{\alpha,\mathbf{k}%
}\left\vert \psi_{j,\mathbf{k}}^{Ab}\right.  \right\rangle ,
\label{eq:TB_projection_coefficients}%
\end{equation}
$c_{j,\alpha}\left(  \mathbf{k}\right)  =\left\langle \Psi_{\alpha,\mathbf{k}%
}\left\vert \psi_{j,\mathbf{k}}^{Ab}\right.  \right\rangle ,$with
the
projection operator%
\begin{equation}
\hat{P}\left(  \mathbf{k}\right)  =\sum_{\alpha}\left\vert
\Psi_{\alpha ,\mathbf{k}}\right\rangle \left\langle
\Psi_{\alpha,\mathbf{k}}\right\vert .
\label{eq:projector}%
\end{equation}
Equation~(\ref{eq:approx_abinit}) contains an approximation of the
\textit{ab-initio} wave functions in so far that the sum over
$\alpha$ extends only over those orbitals that are included in the
tight binding basis
$\mathfrak{B}_{\text{ortho}}$. 
This basis and $\mathfrak{B}_{\text{ortho}}$ of similar ETB\ models
have much fewer basis vectors than the input \textit{ab-initio}
calculation. This rank reduction is a typical outcome of rectangular
transformations such as $\hat{P}$ and is well known in the field of
low rank approximations~\cite{Lang_LRA}.

\textbf{Step 4:} Here, the quality of the ETB fitting is assessed.
In this step, the band structures of the current ETB model
$\varepsilon_{j}^{TB}\left(  \mathbf{k}\right)  $ and the
\textit{ab-initio} input $\varepsilon_{j}^{Ab}\left(
\mathbf{k}\right)  $ are compared. If these sufficiently agree, the
phases of the ETB wave functions are modulated to agree with the
\textit{ab-initio} ones and both wave functions are compared after
that. The ETB Hamiltonian of step 2 is diagonalized in the basis $\mathfrak{B}%
_{\text{ortho}}$ of step 3 to obtain ETB band structures $\varepsilon_{j}%
^{TB}\left(  \mathbf{k}\right)  $ and eigen vectors $\left\vert \psi
_{j,\mathbf{k}}^{TB}\right\rangle $%
\begin{equation}
\hat{H}^{TB}\left(  \mathbf{k}\right)  \left\vert \psi_{j,\mathbf{k}}%
^{TB}\right\rangle =\varepsilon_{j}^{TB}\left(  \mathbf{k}\right)
\left\vert
\psi_{j,\mathbf{k}}^{TB}\right\rangle , \label{eq:TB_eigenfunc}%
\end{equation}
with
\begin{equation}
\left\vert \psi_{j,\mathbf{k}}^{TB}\right\rangle
=\sum_{\alpha}d_{j,\alpha }\left(  \mathbf{k}\right)  \left\vert
\Psi_{\alpha,\mathbf{k}}\right\rangle .
\label{eq:TB_eigenstates_1stNNs}%
\end{equation}

To assess the quality of the ETB results is assessed, different
fitness functions $F_{\varepsilon}$,$F_{m}$ and $F_{\psi}$ are
defined for energies, masses and wave functions respectively. The
$F_{\varepsilon}$ and $F_{m}$ are given by
\begin{eqnarray}
   F_{\varepsilon} &= & \sum_{j,\mathbf{k}}w_{j}^{\varepsilon}\left(  \mathbf{k}%
\right)  \left\vert \varepsilon_{j}^{TB}\left(  \mathbf{k}\right)
-\varepsilon_{j}^{Ab}\left(  \mathbf{k}\right)  \right\vert ^{2}. \\
   F_{m} &=& \sum_{m}w_{m}\left\vert
\frac{m^{Ab}-m^{TB}}{m^{Ab}}\right\vert ^{2}.
\end{eqnarray}
where $w_{j}^{\varepsilon}\left(  \mathbf{k}\right)$ and $w_{m}$ are
weights defined for each target.

As a convention for wave functions phases, another set of ETB\ wave
functions $\left\vert \tilde{\psi}_{j,\mathbf{k}}^{TB}\right\rangle
$ is
introduced%
\begin{equation}
\left\vert \tilde{\psi}_{j,\mathbf{k}}^{TB}\right\rangle =\sum_{i}%
V_{j,i}\left(  \mathbf{k}\right)  \left\vert \psi_{i,\mathbf{k}}%
^{TB}\right\rangle . \label{eq:TB_transformed_WF}%
\end{equation}
The unitary transformation $\hat{V}\left(  \mathbf{k}\right)  $ is
defined by
\begin{equation}
V_{j,i}\left(  \mathbf{k}\right)  =\frac{\left\langle \psi_{j,\mathbf{k}}%
^{TB}\left\vert \psi_{i,\mathbf{k}}^{Ab}\right.  \right\rangle }%
{\lambda\left(  \mathbf{k}\right)  },
\end{equation}
with
\begin{equation}
\lambda\left(  \mathbf{k}\right)
=\sqrt{\frac{1}{N}\sum_{q,p}\left\vert
\left\langle \psi_{q,\mathbf{k}}^{Ab}\left\vert \psi_{p,\mathbf{k}}%
^{TB}\right.  \right\rangle \right\vert ^{2}}.
\end{equation}
Here, the sum over $p$ and $q$ runs over all $N$ ETB states
$\left\vert \psi_{p,\mathbf{k}}^{TB}\right\rangle $ and $N$
\textit{ab-initio }states $\left\langle
\psi_{q,\mathbf{k}}^{Ab}\right\vert $ with equivalent energies
$\varepsilon_{p}^{TB}\left(  \mathbf{k}\right)  \approx\varepsilon_{q}%
^{Ab}\left(  \mathbf{k}\right)  $. With this transformation, the
equation holds
\begin{equation}
\left\langle \psi_{i,\mathbf{k}}^{Ab}\left\vert \tilde{\psi}_{j,\mathbf{k}%
}^{TB}\right.  \right\rangle =\lambda\left(  \mathbf{k}\right)  ,
\end{equation}
for equivalent states. This phase adaption can only work if the ETB\
band structure is close enough to the \textit{ab-initio}
result. 
The ETB wave function fittness is given by%
\begin{equation}
F_{\psi}={\sum_{j,\mathbf{k}}w_{j}^{\psi}}\left(  \mathbf{k}\right)
\left\Vert \left\vert \psi_{j,\mathbf{k}}^{Ab}\right\rangle
-\left\vert \tilde{\psi}_{j,\mathbf{k}}^{TB}\right\rangle
\right\Vert ^{2}{.}
\label{eq:difference_wf}%
\end{equation}
The weights $w_{j}^{\psi}\left(
\mathbf{k}\right)  $ are varying depending on respective fitting
focusses. Deviations of $\left\vert
\tilde{\psi}_{\nu,\mathbf{k}}^{TB}\right\rangle $ from $\left\vert
\psi_{\nu,\mathbf{k}}^{Ab}\right\rangle $ have in general two
reasons: inadequate basis functions and/or eigenfunctions of a
poorly approximated ETB Hamiltonian. Therefore, $F_{\psi}$ can be
estimated as
\begin{align}
\left\Vert \left\vert \psi_{j,\mathbf{k}}^{Ab}\right\rangle
-\left\vert \tilde{\psi}_{j,\mathbf{k}}^{TB}\right\rangle
\right\Vert ^{2}  & \leq2\left\Vert \left[  \hat{I}-\hat{P}\left(
\mathbf{k}\right)  \right]
\left\vert \psi_{j,\mathbf{k}}^{Ab}\right\rangle \right\Vert ^{2}\nonumber\\
&  +2\left\Vert \hat{P}\left(  \mathbf{k}\right)  \left\vert \psi
_{j,\mathbf{k}}^{Ab}\right\rangle -\left\vert \tilde{\psi}_{j,\mathbf{k}}%
^{TB}\right\rangle \right\Vert ^{2}. \label{eq:inequality}%
\end{align}
The first right hand side term of the last equation describes the
deviation of the low-rank approximated \textit{ab-initio} wave
functions. This becomes obvious with the projector property
$\hat{P}^{2}\left(  \mathbf{k}\right) =\hat{P}\left(
\mathbf{k}\right)  $
\begin{equation}
\left\Vert \left[  \hat{I}-\hat{P}\left(  \mathbf{k}\right)  \right]
\left\vert \psi_{j,\mathbf{k}}^{Ab}\right\rangle \right\Vert
^{2}=\left\langle \psi_{j,\mathbf{k}}^{Ab}\left\vert \left[
\hat{I}-\hat{P}\left( \mathbf{k}\right)  \right]  \right\vert
\psi_{j,\mathbf{k}}^{Ab}\right\rangle .
\end{equation}
The second term on the right hand side of Eq.~(\ref{eq:inequality})
contains information about the quality of the eigenfunctions of the
approximate ETB\ Hamiltonian $\hat{H}^{TB}\left(  \mathbf{k}\right)
$. This is understandable when Eqs.~(\ref{eq:approx_abinit}) and
(\ref{eq:TB_transformed_WF}) are inserted into this term%
\begin{eqnarray}
\nonumber   & \left\Vert \hat{P}\left(  \mathbf{k}\right)  \left\vert \psi_{j,\mathbf{k}%
}^{Ab}\right\rangle -\left\vert
\tilde{\psi}_{j,\mathbf{k}}^{TB}\right\rangle
\right\Vert ^{2} =   \\
   &  2-2\operatorname{Re}\left[  \sum_{\alpha,i}c_{j,\alpha}%
^{\dag}\left(  \mathbf{k}\right)  V_{j,i}\left(  \mathbf{k}\right)
d_{i,\alpha}\left(  \mathbf{k}\right)  \right]  .
\end{eqnarray}
The fitness function $F_{\psi}$ represents the major improvement
over the traditional ETB\ eigenvalue fitting (e.g. typically limited
to energies and effective masses). All fitness functions are
minimized by iterating over the steps 3 and 4: the Slater Koster
type parameters for the ETB Hamiltonian $\hat{H}^{TB}\left(
\mathbf{k}\right)  $ and the parameters of the radial ETB\ basis
functions $R_{a,n,l}\left(  r\right)  $ are adjusted for every
iteration of step3.

\textbf{Step 5:} Once the fitness functions are small enough to
cease the iterations, it is assumed that those eigenfunctions of the
ETB\ Hamiltonian $\hat{H}^{TB}\left(  \mathbf{k}\right)  $ that were
subject to the fitting are identical to the eigenfunctions of the
\textit{ab-initio }Hamiltonian $\hat {H}^{Ab}\left(
\mathbf{k}\right)  $ after a transformation $\hat
{A}\left(  \mathbf{k}\right)  $
\begin{equation}
\left\vert \psi_{j,\mathbf{k}}^{TB}\right\rangle
\approx\sum_{i}A_{j,i}\left( \mathbf{k}\right)  \left\vert
\psi_{i,\mathbf{k}}^{Ab}\right\rangle
.\label{eq:A}%
\end{equation}
This transformation $\hat{A}$ is determined by a singular value
decomposition of the rectangular overlap matrix of \textit{ab-initio
}eigenstates with ETB\ eigenstates%
\begin{equation}
\left\langle \psi_{i,\mathbf{k}}^{Ab}\left\vert \psi_{j,\mathbf{k}}%
^{TB}\right.  \right\rangle =\sum_{p}U_{i,p}\left( \mathbf{k}\right)
\Sigma_{p,p}\left(  \mathbf{k}\right) W_{p,j}\left(
\mathbf{k}\right)  .
\end{equation}
The row index $i$ runs over all \textit{ab-initio }eigenstates -
exceeding those that served as fitting targets, whereas the column
index $j$ covers all the ETB eigenfunctions. The $\Sigma$ and $W$ are
square and $U$ is a rectangular matrix. The transformation $\hat{A}$
is then defined as
\begin{equation}
A_{j,i}\left(  \mathbf{k}\right)  =\sum_{p}W_{j,p}\left(  \mathbf{k}%
\right)  U_{p,i}^{\dag}\left(  \mathbf{k}\right)  .
\end{equation}
$\hat{A}$ is constructed from relevant columns of a unitary
transformation. Combining Eqs.~(\ref{eq:A}) and\
(\ref{eq:TB_eigenstates_1stNNs}) allows to
determine the Bloch periodic final basis functions%
\begin{equation}
\left\vert \Psi_{\alpha,\mathbf{k}}^{\text{final}}\right\rangle
=\sum _{i,j}d_{\alpha,j}^{\dag}\left(  \mathbf{k}\right)
A_{j,i}\left( \mathbf{k}\right)  \left\vert
\psi_{i,\mathbf{k}}^{Ab}\right\rangle
.\label{eq:TB_eigenstates_1stNNs_exact}%
\end{equation}
\newline The real space counterpart of $\left\vert \Psi_{\alpha,\mathbf{k}%
}^{\text{final}}\right\rangle $ is given by%
\begin{equation}
\Psi_{\alpha}^{\text{final}}\left(
\mathbf{r}-\mathbf{R}-\mathbf{\tau
}\right)  =\frac{V}{\left(  2\pi\right)  ^{3}}\int_{\textrm{BZ}}\mathrm{d}%
\mathbf{k} e^{ -i\mathbf{k}\cdot\left( \mathbf{R}+\mathbf{\tau
}\right) } \Psi_{\alpha,\mathbf{k}}^{\text{final}}\left(
\mathbf{r}\right)  .\label{eq:real_space_TB_Basis}%
\end{equation}
%
%
\begin{figure*}[ptb]
\includegraphics[width=1.8\columnwidth]{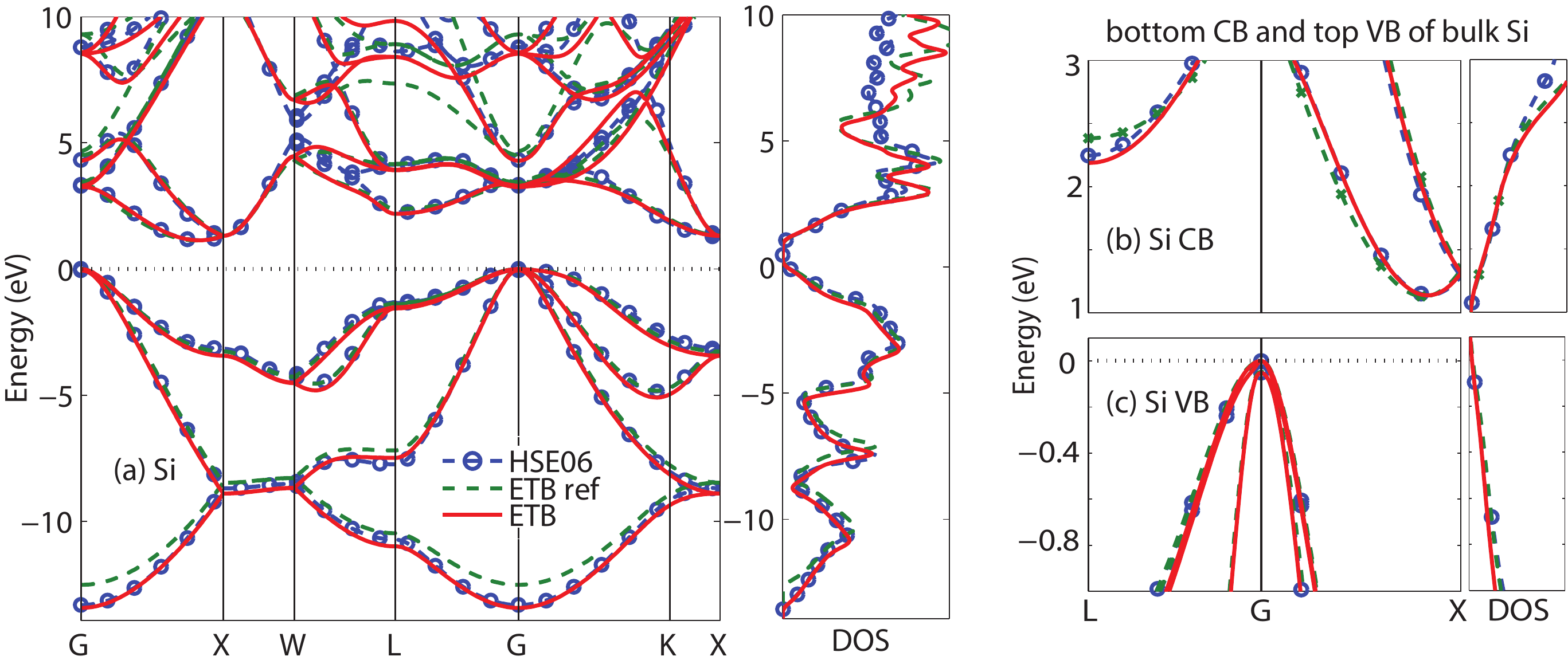}\newline%
\caption{Band structure and density of states of bulk Si. ETB band
structure agree with the HSE06 band structure (a),
especially for bottom conduction bands (b) and top valence bands(c) around Fermi level. }%
\label{fig:Si Ek and DOS}%
\end{figure*}
\section{Results\label{sec:results}}
In this work, \textit{ab-initio }level calculations of Si and GaAs
systems were performed with VASP \cite{VASP_Kresse}. The HSE06
hybrid functional \cite{Vasp_HSE} is used to produce reasonable band
gaps in both the bulk and the UTB cases. In all HSE06 calculations,
a cutoff energy of 350eV is used. $\Gamma$-point centered Monkhorst
Pack kspace grids are used for both bulk and UTB systems. The size
of the kspace grid for bulk calculations is a $6\times6\times6$,
while one for UTB is $6\times6\times1$. k-points with integration
weights equal to zero are added to the original $6\times6\times6$ or
$6\times6\times1$ grids in order to generate energy bands with
higher k-space resolution. The spin orbit coupling is included in
band structure calculations. Small hydrostatic strains up to $0.3\%$
are introduced to adjust the bulk band gaps in order to match
experimental results. The lattice const used in this work is given
by table \ref{table:SKtype_tb_parameters}.
\begin{figure*}[ptb]
\includegraphics[width=1.8\columnwidth]{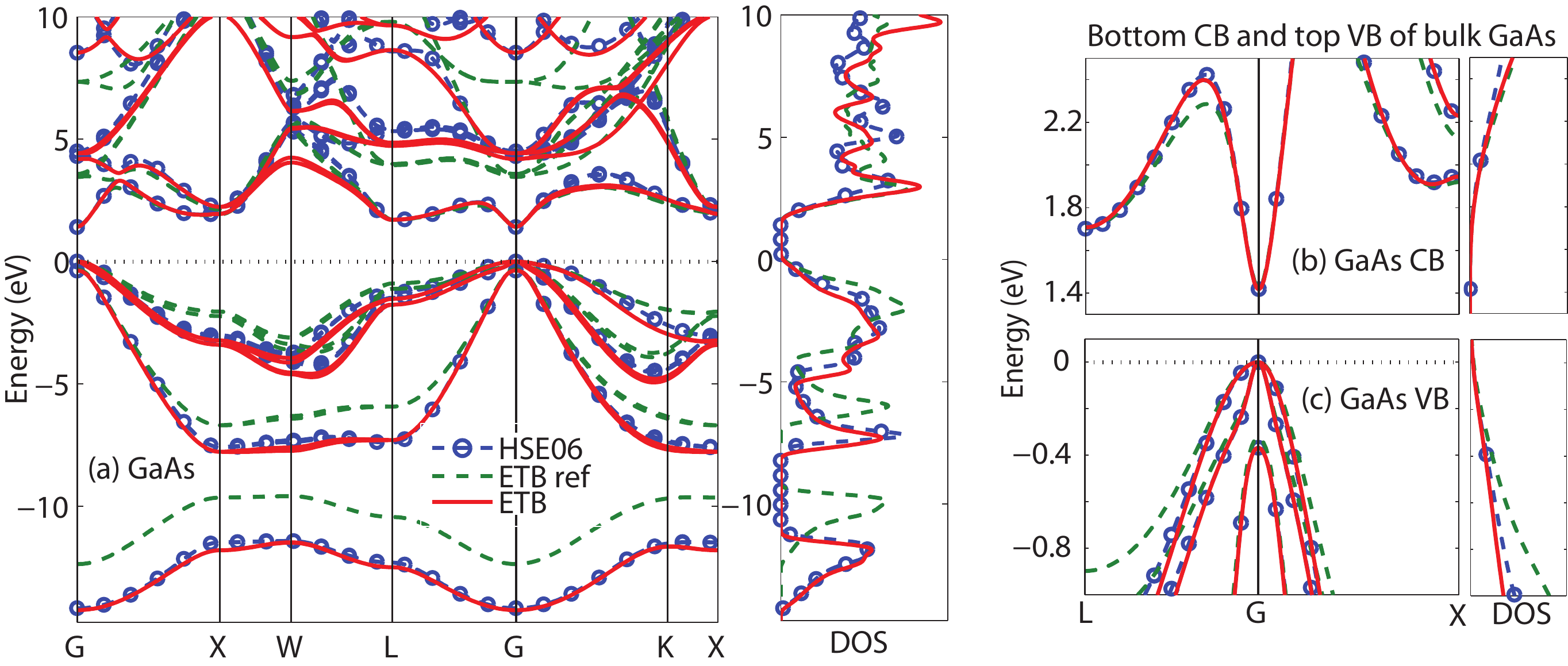}\newline%
\caption{Band structure and density of states of bulk GaAs. ETB band
structure agree with the HSE06 band structure (a),
especially for bottom conduction bands (b) and top valence bands(c) around Fermi level.}%
\label{fig:GaAs Ek and DOS}%
\end{figure*}

\subsection{Application to Bulk Materials\label{sec:application_to_bulk_materials}}
\begin{table}[ptb]
\centering {\small
\begin{tabular}
[c]{c||c}\hline
Si & GaAs\\\hline%
\begin{tabular}
[c]{cc}\hline
$a_0$ & $5.43 \AA $ \\
                     $E_{s }$ &           $-2.803316$ \\
                     $E_{p }$ &           $4.096984$ \\
                    $E_{s^* }$ &          $25.163115$ \\
                     $E_{d }$ &          $12.568228$ \\
                 $\Delta$ &           $0.021926$ \\
                 & \\
          $V_{s s \sigma}$ &          $-2.066560$ \\
       $V_{s^* s^* \sigma }$ &          $-4.733506$ \\
        $V_{s s^* \sigma }$ &          $-1.703630$ \\
         $V_{s p \sigma }$ &           $3.144266$ \\
        $V_{s^* p \sigma }$ &           $2.928749$ \\
         $V_{s d \sigma }$ &          $-2.131451$ \\
        $V_{s^* d \sigma }$ &          $-0.176671$ \\
        $V_{p p  \sigma }$ &           $4.122363$ \\
            $V_{p p \pi }$ &          $-1.522175$ \\
        $V_{p d \sigma  }$ &          $-1.127068$ \\
            $V_{p d \pi }$ &           $2.383978$ \\
        $V_{d d \sigma  }$ &          $-1.408578$ \\
            $V_{d d \pi }$ &           $2.284472$ \\
         $V_{d d \delta }$ &          $-1.541821$ \\
\end{tabular}
&
\begin{tabular}
[c]{cc|cc}\hline $a_0$ & $5.6307 \AA $ &  & \\
$E_{s_a}$ &           $-8.063758$ &                     $E_{s_c}$ &           $-1.603222$                      \\
$E_{p_a}$ &           $3.126841$  &                     $E_{p_c}$ &           $4.745896$                       \\
$E_{s^*_a}$ &           $21.930865$&                    $E_{s^*_c}$ &          $23.630466$                     \\
$E_{d_a}$ &          $13.140998$  &                     $E_{d_c}$  &          $14.807586$                       \\
$\Delta_{a}$ &         $0.194174$&                     $\Delta_{c}$ &       $0.036594$                  \\
&  &  & \\
          $V_{s_a s_c \sigma}$ &          $-1.798514$ &   & \\
       $V_{s^*_a s^*_c \sigma }$ &          $-4.112848$ &   & \\
        $V_{s_a s^*_c \sigma }$ &          $-1.258382$ &  $V_{s_c s^*_a \sigma }$ &          $-1.688128$ \\
         $V_{s_a p_c \sigma }$ &           $3.116745$ &  $V_{s_c p_a \sigma }$ &           $2.776805$  \\
        $V_{s^*_a p_c \sigma }$ &           $1.635158$ &  $V_{s^*_c p_a \sigma }$ &           $3.381868$ \\
         $V_{s_a d_c \sigma }$ &          $-0.396407$ &  $V_{s_c d_a \sigma }$ &          $-2.151852$\\
        $V_{s^*_a d_c \sigma }$ &          $-0.145161$ &  $V_{s^*_c d_a \sigma }$ &          $-0.810997$\\
        $V_{p_a p_c  \sigma }$ &           $4.034685$ &  & \\
            $V_{p_a p_c \pi }$ &          $-1.275446$ &  & \\
        $V_{p_a d_c \sigma  }$ &          $-1.478036$ & $V_{p_c d_a \sigma  }$ &          $-0.064809$ \\
            $V_{p_a d_c \pi }$ &           $1.830852$ & $V_{p_c d_a \pi }$ &           $2.829426$ \\
        $V_{d_a d_c \sigma  }$ &          $-1.216390$ &  & \\
            $V_{d_a d_c \pi }$ &           $2.042009$ &  & \\
         $V_{d_a d_c \delta }$ &          $-1.829113$ &  & \\
\end{tabular}
\\\hline
$H_{Si}$ & $H_{c}\quad \textrm{and}\quad H_{a}$
\\\hline%
\begin{tabular}
[c]{cc}\hline
$E_{s_{H} }$ & $-3.056510$\\
$V_{s_{H} s_{Si} \sigma}$ & $-4.859509$\\
$V_{s_{H} p_{Si} \sigma}$ & $3.776178$\\
$V_{s_{H} s^{*}_{Si} \sigma}$ & $0.0$\\
$V_{s_{H} d_{Si} \sigma}$ & $-0.007703$\\
$\delta_{Si}$ & $-0.276789$\\\hline
\end{tabular}
&
\begin{tabular}
[c]{cc|cc}\hline
$E_{s_{H_c} }$ & $2.758428$ & $E_{s_{H_a} }$ & $-0.308397$  \\
$V_{s_{H_c} s_{a}\sigma}$ & $-2.960420$ & $V_{s_{H_a} s_{c}\sigma}$ & $-3.151427$  \\
$V_{s_{H_c} p_{a} \sigma}$ & $5.490764$ & $V_{s_{H_a} p_{c} \sigma}$ & $3.539284$  \\
$V_{s_{H_c} s^{*}_{a} \sigma}$ & $0.0$ & $V_{s_{H_a} s^{*}_{c}\sigma}$ & $-0.129904$  \\
$V_{s_{H_c} d_{a}\sigma}$ &$-1.727690$ & $V_{s_{H_a} d_{c} \sigma}$ & $-0.252733$  \\
$\delta_{a}$ & $-0.266815$ & $\delta_{c}$ & $-0.586952$  \\\hline
\end{tabular}
 \\\hline%
\end{tabular}
}\caption{Slater Koster type ETB parameters of bulk Si and GaAs, and passivation parameters of UTBs.  }%
\label{table:SKtype_tb_parameters}%
\end{table}
For bulk Si and GaAs, fitting targets include the band structures of
the lowest 16 bands (with spin degeneracy) along high symmetry
directions, important effective masses and wave functions at high
symmetry points such as $\Gamma$, $L$ and $X$ points. ETB basis
functions in real space is reconstructed on $6\times6\times6$
$\Gamma$ center k space grid using Eq
(\ref{eq:real_space_TB_Basis}).

\begin{table}[h]
\begin{tabular}
[c]{c|c}\hline
& Si\\\hline%
\begin{tabular}
[c]{c}%
targets\\\hline
$E_{g}(\Gamma)$\\
$E_{g}(X)$\\
$E_{g}(L)$\\
$\Delta_{SO}$\\
\\
$m_{hh100}$\\
$m_{hh110}$\\
$m_{hh111}$\\
$m_{lh100}$\\
$m_{lh110}$\\
$m_{lh111}$\\
$m_{so100}$\\
$m_{so110}$\\
$m_{so111}$\\
\\
$m_{cXl}$\\
$m_{cXt}$\\
\end{tabular}
&
\begin{tabular}
[c]{cccc}%
TB Ref & HSE06 & TB & error $(\%)$ \\\hline
              $3.399$ &              $3.302$ &              $3.244$ &                $1.8$ \\
              $1.131$ &              $1.142$ &              $1.139$ &                $0.2$ \\
              $2.383$ &              $2.247$ &              $2.188$ &                $2.6$ \\
              $0.047$ &              $0.051$ &              $0.052$ &                $0.8$ \\
&  &  & \\
              $0.299$ &              $0.281$ &              $0.282$ &              $0.097$ \\
              $0.633$ &              $0.566$ &              $0.572$ &              $0.977$ \\
              $0.796$ &              $0.704$ &              $0.714$ &              $1.433$ \\
              $0.232$ &              $0.206$ &              $0.204$ &              $1.001$ \\
              $0.165$ &              $0.151$ &              $0.149$ &              $0.937$ \\
              $0.156$ &              $0.143$ &              $0.142$ &              $0.927$ \\
              $0.266$ &              $0.244$ &              $0.242$ &              $0.809$ \\
              $0.266$ &              $0.244$ &              $0.242$ &              $0.795$ \\
             $0.267$ &              $0.244$ &              $0.242$ &              $0.770$ \\
&  &  & \\
              $0.887$ &              $0.928$ &              $0.857$ &              $7.615$ \\
              $0.225$ &              $0.207$ &              $0.215$ &              $3.544$ \\
\end{tabular}
\\\hline
\end{tabular}
\centering \caption{Targets comparison of bulk Si. Critical band
edges and effective masses
at $\Gamma$, $X$ and $L$ points by ETB and HSE06 calculations are compared.}%
\label{tab:targets_comparison_Si}%
\end{table}
 The band structures and DOS of bulk Si and GaAs
(HSE06 vs ETB) are shown in Fig.~\ref{fig:Si Ek and DOS} and
\ref{fig:GaAs Ek and DOS} respectively. The band structures using
existing Si and GaAs ETB
parameters\cite{Boykin_SiGe,Boykin_TB_strain} are also shown in
corresponding figures. The ETB band structures and DOS using
parameters generated by this work show better agreement with the
corresponding hybrid functional results compared with the existing
parameterizations. For bulk Si, the existing parameterization shows
a unexpected low $s^{*}$ band around 5 eV above topmost valence
bands. In the traditional fitting process, the $s^{*}$ band shows a
strong preference for moving downward~\cite{Boykin_SiGe}. Due to
large number of parameters to be determined, traditional (energy-gap
and effective-mass) based fitting procedures can find local minima
in their fitness functions corresponding to wave functions
significantly different from those predicted by \textit{ab-initio}
methods. The present method has the important advantage that
optimization involves not only masses and gaps but also
wavefunctions. Thus the ETB wavefunctions can be kept close to their
\textit{ab-initio} counterparts. For GaAs, the existing
parameterization shows 2 eV higher $s$-type low lying valence bands.
The ETB parameters of bulk Si and GaAs are listed in table
\ref{table:SKtype_tb_parameters}. It can be seen from tables
\ref{tab:targets_comparison_Si} and
\ref{tab:targets_comparison_GaAs}, the anisotropic hole masses by
ETB show a remarkable agreement with HSE06 results. The principal
authors of the previous works\cite{Boykin_TB_strain,Boykin_SiGe}
explicitly pointed out that fitting hole masses had been very
difficult with the previous methods.
\begin{table}[h]
\begin{tabular}
[c]{c|c}\hline
& GaAs\\\hline%
\begin{tabular}
[c]{c}%
targets\\\hline
$E_{g}(\Gamma)$\\
$E_{g}(X)$\\
$E_{g}(L)$\\
$\Delta_{SO}$\\
\\
$m_{hh100}$\\
$m_{hh110}$\\
$m_{hh111}$\\
$m_{lh100}$\\
$m_{lh110}$\\
$m_{lh111}$\\
$m_{so100}$\\
$m_{so110}$\\
$m_{so111}$\\
\\
$m_{c100}$\\
$m_{c110}$\\
$m_{c111}$\\
$m_{cXl}$\\
$m_{cXt}$\\
$m_{cLl}$\\
$m_{cLt}$\\
\end{tabular}
&
\begin{tabular}
[c]{cccc}%
TB Ref & HSE06 & TB & error$(\%)$\\\hline
              $1.424$ &              $1.418$ &              $1.416$ &                $0.2$ \\
              $1.900$ &              $1.919$ &              $1.910$ &                $0.5$ \\
              $1.707$ &              $1.702$ &              $1.708$ &                $0.3$ \\
              $0.326$ &              $0.368$ &              $0.367$ &                $0.1$ \\
&  &  & \\
              $0.383$ &              $0.310$ &              $0.337$ &              $8.510$ \\
              $0.667$ &              $0.573$ &              $0.619$ &              $7.879$ \\
              $0.853$ &              $0.750$ &              $0.813$ &              $8.507$ \\
              $0.085$ &              $0.082$ &              $0.083$ &              $0.744$ \\
              $0.078$ &              $0.073$ &              $0.074$ &              $1.614$ \\
              $0.076$ &              $0.071$ &              $0.072$ &              $1.715$ \\
              $0.166$ &              $0.164$ &              $0.160$ &              $1.998$ \\
              $0.166$ &              $0.164$ &              $0.160$ &              $2.037$ \\
              $0.166$ &              $0.164$ &              $0.160$ &              $2.041$ \\
                   & & & \\
              $0.068$ &              $0.065$ &              $0.067$ &              $2.787$ \\
              $0.068$ &              $0.066$ &              $0.067$ &              $2.790$ \\
              $0.068$ &              $0.065$ &              $0.067$ &              $2.781$ \\
              $1.526$ &              $1.577$ &              $1.480$ &              $6.142$ \\
              $0.177$ &              $0.215$ &              $0.204$ &              $5.083$ \\
              $1.743$ &              $1.626$ &              $1.446$ &             $11.055$ \\
              $0.099$ &              $0.111$ &              $0.136$ &             $22.614$ \\
\end{tabular}
\\\hline
\end{tabular}
\centering \caption{Targets comparison of bulk GaAs. Critical band
edges and effective
masses at $\Gamma$, $X$ and $L$ from TB and HSE06 calculations are compared.}%
\label{tab:targets_comparison_GaAs}%
\end{table}

\begin{figure*}[ptb]
\centering
\includegraphics[width=1.7\columnwidth]{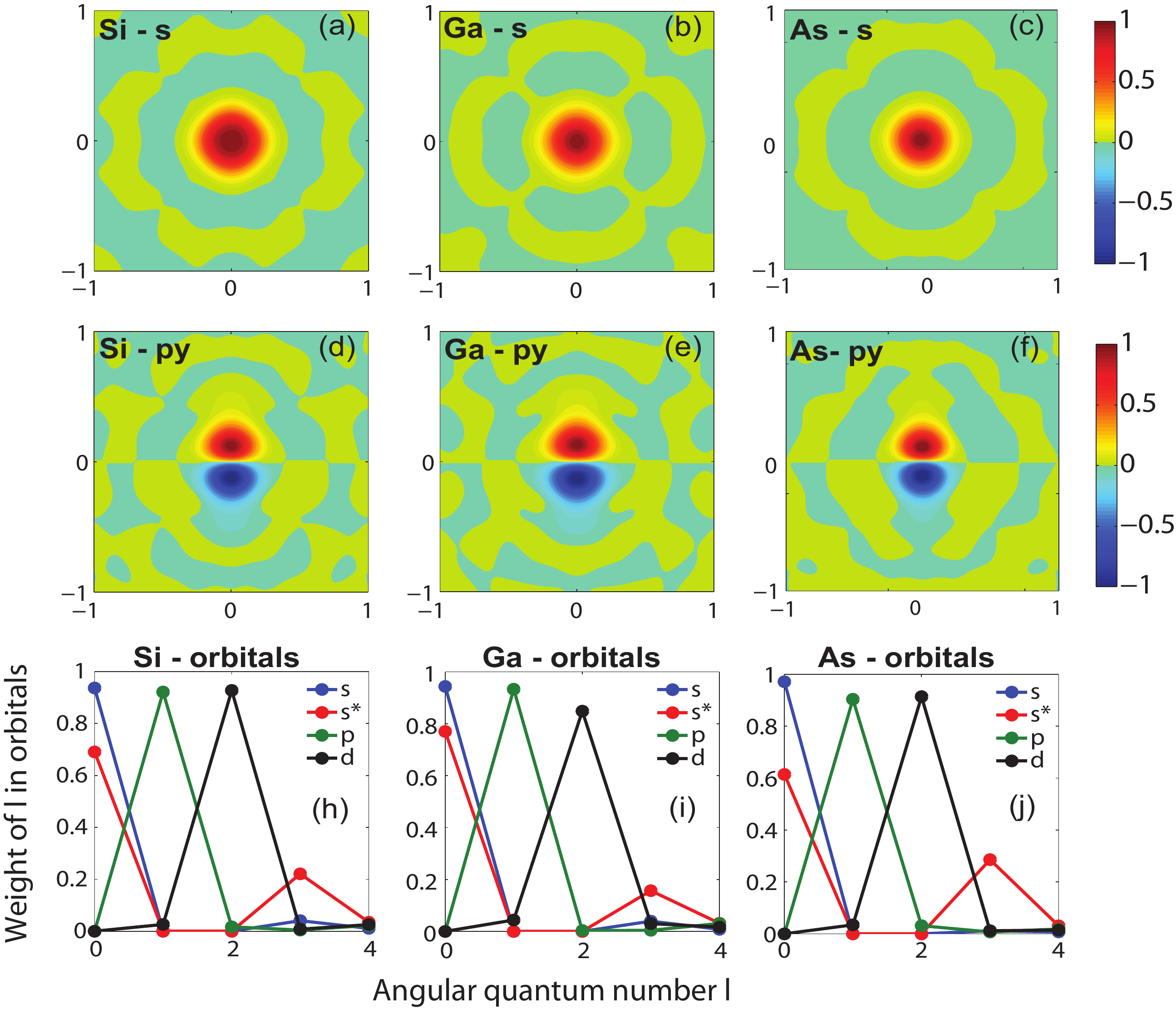}\caption{
Contours of selected ETB basis functions of Si((a),(d)), Ga ((b),(e))
and As ((c),(f)) atoms. (a),(b) and (c) correspond to the contours
of $s$ orbitals of Si, Ga and As atoms in x-y plane. (d),(e) and (f)
 correspond to the contours of $p_{y}$ orbitals of Si, Ga, and As atoms. (g),(h)
and (i) show the contribution of different angular momentums in
basis functions of Si ,Ga, and As atoms. The ETB basis functions of Si and GaAs
are highly localized basis functions with one dominant angular momentum.}%
\label{fig_basisfn_Si_Ga_As}%
\end{figure*}
The orthogonal ETB basis functions $\mathfrak{B}_{\text{final}}$ of
Si, Ga and As atoms are shown in Fig.~\ref{fig_basisfn_Si_Ga_As}.
The ETB basis functions are slightly environment dependent because
they are orthogonal. Thus the ETB basis functions are not invariant
under arbitrary rotations but invariant under symmetry operations
within $T_{d}$ group, as pointed out by Slater and
Koster~\cite{Slater_Tightbinding}. It can be seen from
Fig.~\ref{fig_basisfn_Si_Ga_As}.(a) to (f) that the $s$ and $p$
orbitals show $s$ and $p$ features near the atom. More complicated
patterns in the area further away from the atom can be observed.
These complicated patterns correspond to components with high
angular momentums. The feature of orthogonal ETB basis function
resembles the augmented basis functions used in \textit{ab-initio}
level calculations such as Augmented Plane Waves(APWs) and Muffin
Tin Orbitals(MTOs). The orthogonal ETB basis functions have multiple
angular parts in each orbital as shown by
Fig.~\ref{fig_basisfn_Si_Ga_As}.(g),(h) and (i). The $s$, $p$ and
$d$ type ETB basis functions are dominated by components with $l=0$,
$1$ and $2$ respectively. More than $90\%$ for the $s$,$p$ and $d$
orbitals are comprised of their $l=0$, $1$ and $2$ components
respectively. The excited $s^{*}$ type ETB basis functions have
higher angular momentum and the $l = 0$ components have
contributions of $60\%$ to $70\%$. The second largest contribution
in $s^{*}$ orbitals is the $f$ component with $l=3$. The $f$
component attached to the $s^{*}$ orbitals have angular part
equivalent to real space function $xyz$. This is a result of the
existence of $xyz$-like crystal field near each atom in zincblende
and diamond structures. 

\subsection{Application to UTBs\label{sec:application_to_UTB}}
\begin{figure*}[ptb] \centering
\includegraphics[width=1.5\columnwidth]{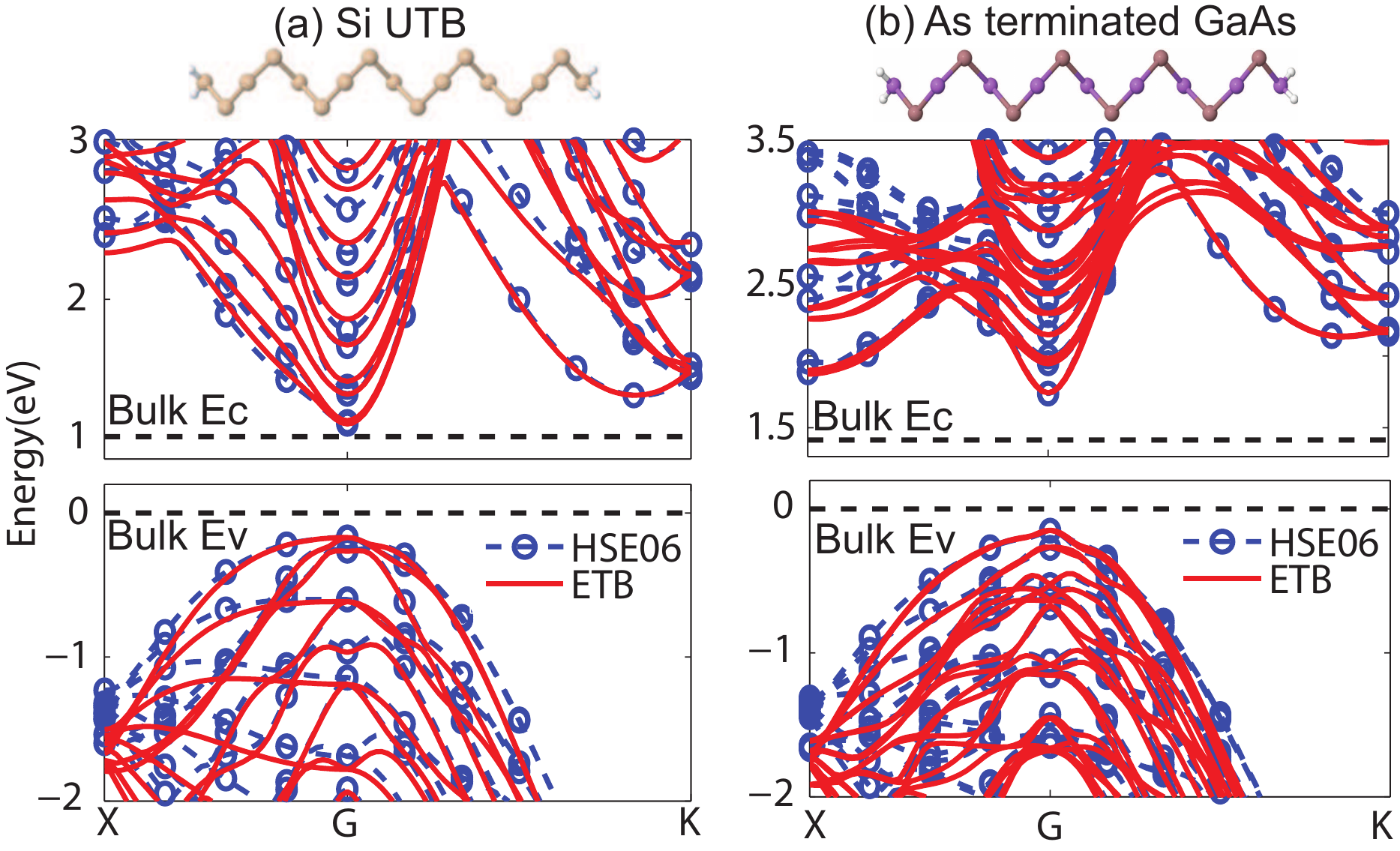}
\caption{ Band structures of 001 Si (a) and As terminated GaAs (b)
UTBs by ETB agree with HSE06 band structures, demonstrating the bulk
Si and GaAs ETB parameters are transferable to UTB cases. All UTBs
contain 17 non-Hydrogen atomic layers(with thickness 4$a_{0}$).}
\label{fig_UTB_Ek}
\end{figure*}
To validate the transferability of the ETB model, band structures and
eigen functions of [001] UTBs passivated by Hydrogen atoms are
calculated by both HSE06 and ETB models.  The current calculations
assume no strain in the UTBs. In the HSE06 calculations, charged
hydrogen atoms are used to passivate the dangling bonds of the
surface atoms in GaAs UTBs. The surface As and Ga atoms are
passivated by charged hydrogen atoms with 3/4 ( denoted by $H_c$ )
and 5/4 ( denoted by $H_a$ ) electron respectively. The charged
hydrogen atoms neutralize most of the surface induced electric field
in the UTBs. As a result, the charge distribution and local
potential shows almost flat envelopes inside the UTBs. Small
deviation of potential can only be observed at the surface Si/Ga/As
atoms. The nearly flat potential envelope suggests geometry
dependent build-in potentials are needed only for surface atoms.
Thus the comparisons between self-consistent hybrid functional
calculations and single shot ETB calculations are fair.

\begin{figure}[ptb] \centering
\includegraphics[width=1\columnwidth]{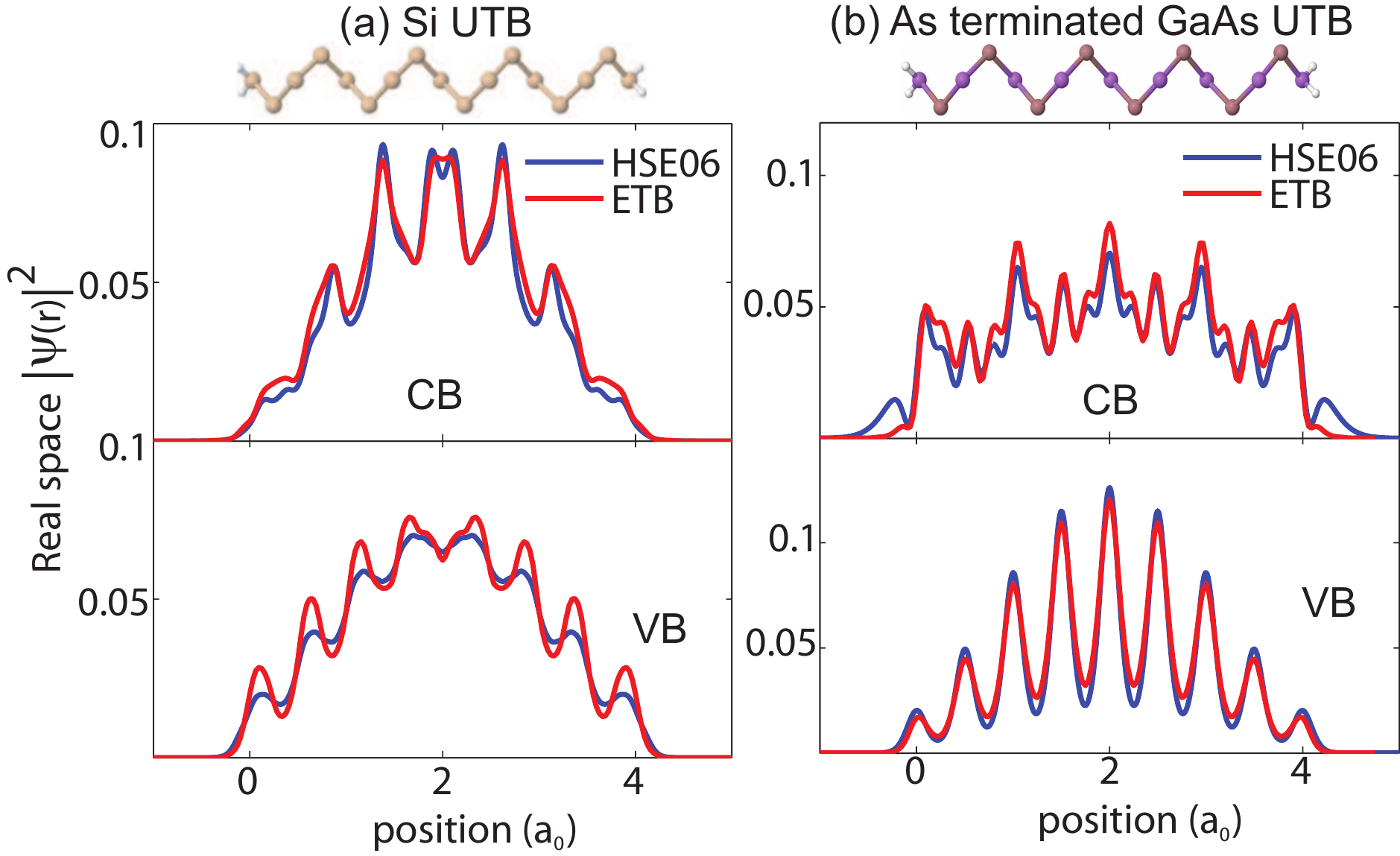}
\caption{Planar averaged real space probability amplitudes of lowest
conduction and topmost valance states of 001
 Si (a) and As terminated GaAs UTBs (b) by HSE06 and ETB calculations.
 With the real space TB basis functions,
 the realspace probability amplitudes of TB calculations show reasonable agreement with the HSE06 probability amplitudes.
 UTBs contain 17 non-Hydrogen atomic layers(with thickness 4$a_{0}$). }
\label{fig_UTB_WF}\end{figure}
The HSE06 calculations show that the
Hydrogen orbitals contribute to the deep valence bands, thus
Hydrogen atoms are considered explicitly into the ETB Hamiltonian of
UTBs in this work. 1s orbital is used as the ETB basis function for
Hydrogen atoms. The explicit passivation model include extra Slater
Koster type ETB parameters $E_{s_{H} s}$, $V_{s_{H} s \sigma}$,
$V_{s_{H} p \sigma}$,$V_{s_{H} s^{*} \sigma}$ and $V_{s_{H} d
\sigma}$. Further more, a geometry and element dependent potential
$\delta$ is included for surface atoms. The onsite energies of the
surface atoms are shifted by $\delta$. The onsite energy of the
surface Ga atoms are thus $E_{\alpha_c} + \delta_{H_c}$; and for
surface As atoms, the onsite energies are $E_{\alpha_a} +
\delta_{H_c}$. Here the $\alpha$ stands for $s$,$p$,$d$ and $s^*$
orbitals.
 ETB parameters of Si/GaAs in Si/GaAs UTBs are identical
with the parameters of unstrained bulk materials provided in section
\ref{sec:application_to_bulk_materials}.

\begin{figure}[ptb]
\centering
\includegraphics[width=\columnwidth]{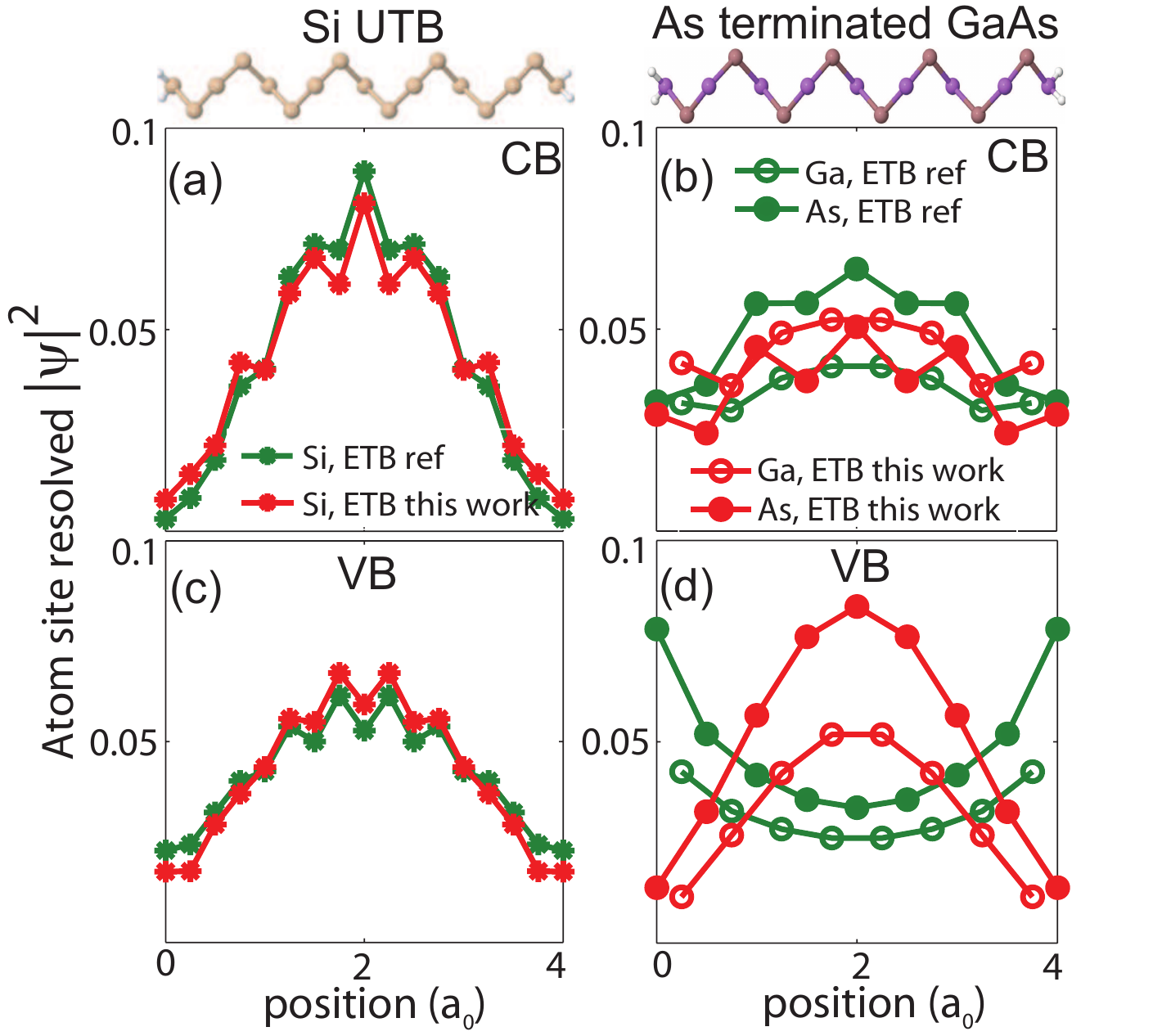}
\caption{ ETB atom site resolved probability amplitudes of Si
((a),(c)), and As terminated GaAs ((b),(d)) UTBs using ETB
parameters in this work and previous
work~\cite{Boykin_SiGe,Boykin_TB_strain}. The ETB atom site
probability using different parameters are qualitatively similar in
Si UTB, while the ETB atom site probability in As GaAs are more
sensitive to the parameter sets and passivation models, i.e. the
valence states with parameters and passivation model by previous
work are not confined. UTBs contain 17 atomic layers(thickness is
4$a_{0}$). }\label{fig_UTB_WF_TB}%
\end{figure}
To determine the ETB parameters of H-passivation, band structures and
real space wave functions of selected bands near the Fermi level of
the UTBs are considered as fitting targets. The inclusion of wave
functions as targets serves the purpose of correcting possible
problematic states. The target Si/GaAs UTBs contain 17 non-Hydrogen
atomic layers. Parameters for Hydrogen atoms are also shown in table
\ref{table:SKtype_tb_parameters}. In GaAs UTBs, As and Ga are
passivated by Hydrogen atoms with different charge, thus the
Hydrogen atoms have different onsite energies when different types
of atoms are passivated. The Hydrogen atoms bonding with As atoms
are charged positively while the ones bonding with Ga atoms are
charged negatively. Consequently, the $H_c$ which forms bond with As
have a higher onsite energy than the $H_a$ which forms bond with Ga.

\begin{figure}[ptb]
\centering
\includegraphics[width=\columnwidth]{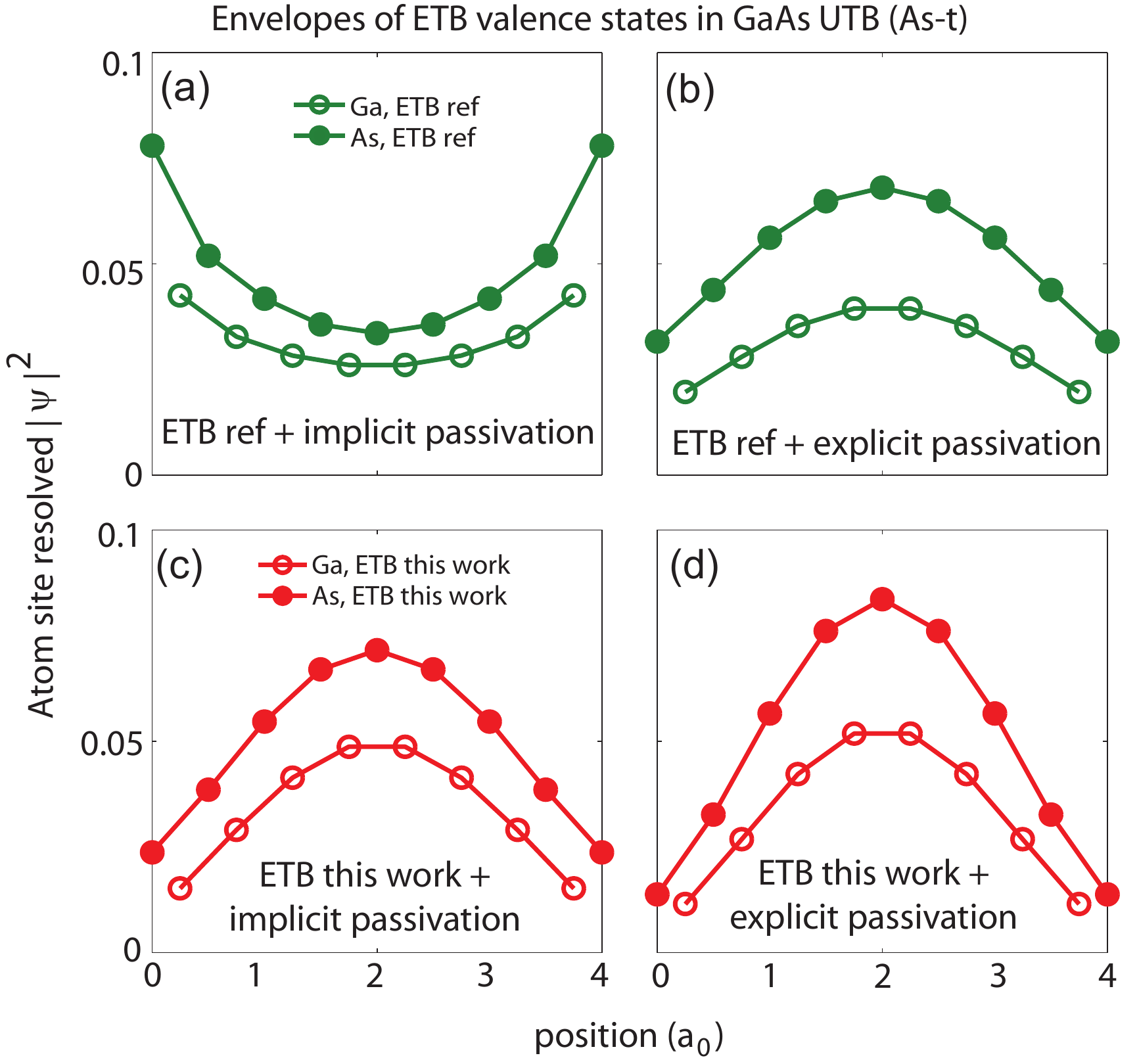}\caption{
Comparison of ETB wave functions using different ETB parameters and
passivation model. (a) and (b) use ETB parameters in ref.~
\onlinecite{Boykin_TB_strain}. (c) and (d) use ETB parameters in
this work. (a) and (c) correspond to implicit passivation model
\cite{TB_passivation}; (b) and (d) correspond to explicit
passivation model. The ETB parameters with the explicit passivation
model shows the most confined states, while the previous parameters
and implicit passivation model lead to less confined states.}
\label{fig_GaAs_wf_para_passivation_cmp}%
\end{figure}
Band structures of Si/GaAs UTBs are shown in Fig.~\ref{fig_UTB_Ek}.
The ETB band structures match the HSE06 band structures well for
energies ranging from 1eV below the topmost valence bands to 1eV
above the lowest conduction bands. Using the explicit ETB basis
functions, ETB wave functions of UTBs with subatomic resolution are
obtained and can be compared with corresponding HSE06 wave
functions. Planar averaged probability amplitudes of wave functions
of the lowest conduction band and top most valence bands in Si/GaAs
UTBs are shown in Fig.~\ref{fig_UTB_WF}. 
It can be seen that not
only the envelope but also details in subatomic resolution of the
ETB planar averaged $|\psi|^2$ show agreement with corresponding
HSE06 results. On the other hand, Fig.~\ref{fig_UTB_WF_TB} compares
the ETB atom site resolved probability amplitudes among ETB models
in present and previous works
(Ref.~\onlinecite{Boykin_SiGe,Boykin_TB_strain}). The cations and
anions in GaAs UTBs form different envelopes for all of the
presented states. The lowest conduction and highest valence states
turn out to be well confined states in Si UTBs in all of the
calculations. While, in GaAs UTBs, the lowest conduction states has
significant contribution from the surface atoms. In Si 
ETB probability amplitudes by previous parametrizations show similar
envelopes compared to the ETB and HSE06 probability amplitudes in
this work. Fig.~\ref{fig_UTB_WF_TB} (d) shows the problematic
valence states in As terminated GaAs UTB by parameters of previous
work. The corresponding valence states by this work turn out to be a
well confined ones. To investigate this issue in more detail, in
Fig.~\ref{fig_GaAs_wf_para_passivation_cmp}, ETB atom site resolved
probability amplitudes for the topmost valence states of the four
possible As-terminated GaAs UTBs are plotted: (a) previous
parameters \cite{Boykin_TB_strain} and implicit passivation
\cite{TB_passivation}; (b) previous parameters and explicit
passivation; (c) new parameters and implicit passivation; (d) new
parameters and explicit passivation. It is clear that, for a given
set of bulk parameters, the implicit passivation model leads to
wavefunctions that are less-confined than those of the explicit
passivation model. On the other hand, with the same passivation
model, the ETB parameters by this works shows more confined top
valence states than the existing ETB parameters. Thus the
un-confined ETB state using the existing parameter set and implicit
passivation model appears to be due to both the bulk GaAs parameters
and the passivation model. The implicit model\cite{TB_passivation}
replaces the $s$- and $p$-orbitals of the surface atoms by sp3
hybrids and raises the energy of the dangling hybrids by
$\delta_{sp3} = 30eV$. The $d$- and $s^*$-orbitals are left
completely un-passivated, and the unconfined states of
Fig.~\ref{fig_GaAs_wf_para_passivation_cmp} (a) are only slightly
affected by changing the value of $\delta_{sp3}$. The impact of
alternate implicit passivation model to explicit passivation model
is obvious by comparing sub-figures (a) to (b), as well as (c) to
(d). To better understand the role of bulk parameters to this
behavior, we experimented by reducing the magnitude of the
nearest-neighbor $p_a$-$d_c$ coupling parameters in both sets as
$V_{p_ad_c\sigma} \rightarrow V_{p_ad_c\sigma} + 0.3eV$,
$V_{p_ad_c\pi} \rightarrow V_{p_ad_c\pi} - 0.3eV$. Remarkably, in
both cases the topmost valence-band state became much more confined.
Bulk valence band wave functions in modified and original parameter
sets tell the story: The general trend is that bulk sets which
generate more $p$-like top of VB states give better confinement
under passivation (and especially implicit passivation) than do
those with higher d-content. The reduction of
$\left|V_{p_ad_c\sigma}\right|$ and $\left|V_{p_ad_c\pi}\right|$
lead to more $p$-like top VB states. Ga terminated case has less
passivation problems because its top-of-VB bulk
states have more contribution from the As atoms than from the Ga atoms.

\begin{figure*}[ptb]
\centering
\includegraphics[width=1.6\columnwidth]{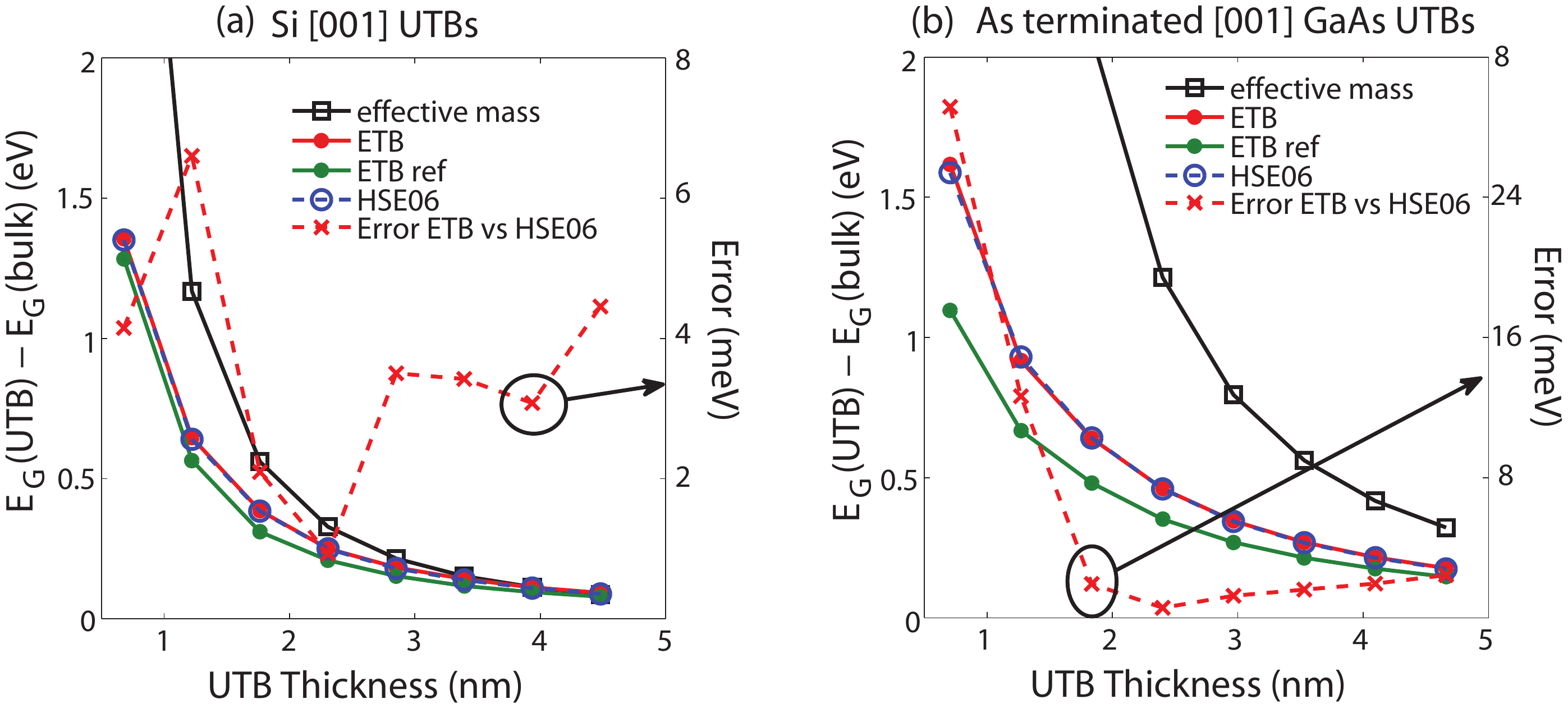}\caption{Band
gaps of Si UTBs (a) and As terminated UTBs (b) by HSE06 and ETB
calculations. For the presented UTBs with thickness ranging from 1nm
to 4.5nm, the ETB band gaps have discrepancies of less than 10meV
compared with HSE06 ones. The band gap changes by effective mass
calculation show agreement with HSE06 for Si UTBs thicker than 3nm.
While the effective mass calculations has obvious discrepancies for
all GaAs UTBs. The HSE06 and ETB calculations using parameters by
this work consider Hydrogen atoms explicitly,
while the ETB calculations using parameters by previous work is based on implicit passivation model\cite{TB_passivation}. }%
\label{fig_UTB_gap}%
\end{figure*}
Fig.~\ref{fig_UTB_gap} shows the band gaps of the Si and GaAs [001]
UTBs as functions of UTB thickness. With the ETB parameters by this
work, the ETB bandgaps of Si and GaAs UTBs with thickness from 0.5nm
to 4nm agree well with the gaps by HSE06 calculations. The ETB
bandgaps of Si UTBs using parameters from previous work also show
good agreement with the HSE06 results. However the ETB bandgaps of
GaAs UTBs using parameters from previous work and implicit
passivation model are of around $20\% $ lower than the Hybrid
functional results. The gaps of GaAs UTBs terminated with Ga and As
atoms are very close in value for both Hybrid functional and ETB
results in this work, however the gaps of GaAs UTBs terminated with
Ga and As atoms by previous parameterizations and implicit
passivation model show 0.1 to 0.2eV discrepancies. The band gap
change in Si UTBs thicker than 3nm can be model by effective mass
model( assuming parabolic E-k relation ). While in the GaAs UTBs,
the discrepancies between effective mass calculations and HSE06 or
TB calculations are obvious for all GaAs UTBs presented, suggesting
the non-parabolic feature of the GaAs valleys have significant
impact to GaAs nano structures. The gaps by previous
parameterization with implicit passivation model of As terminated
GaAs UTBs has lower confined energies due to the unconfined valence
states.

\section{Conclusion\label{Sec:conclusion}}
It has been shown that the existing ETB parameterization together
with the implicit passivation model gives unphysical states in As
terminated GaAs UTB calculations. A more reliable technique of
\textit{ab-initio} mapping which generates ETB parameters and basis
functions from \textit{ab-initio} is developed. The
\textit{ab-initio} mapping process is applied to both bulk Si and
GaAs. Slater Koster type ETB parameters within 1st nearest neighbour
approximation and highly localized ETB basis functions are obtained.
The ETB parameters and basis functions of Si and GaAs are validated
in corresponding UTB systems with passivation models that consider
Hydrogen atom explicitly. Band gaps in Si and GaAs UTBs with
different thickness are also calculated by HSE06, ETB and effective
mass model. Compared with the existing ETB parameterizations and
implicit passivation model, the ETB calculations in this work show
good agreements with HSE06 calculations in both band structures and
wave functions. This work shows that the ETB parameters by
\textit{ab-initio} mapping have good transferability. The mapping
method developed here significantly reduces the uncertainty in both
bulk and passivation models.

\begin{acknowledgments}
nanoHUB.org computational resources operated by the Network for
Computational Nanotechnology funded by NSF are utilized in this
work. Evan Wilson from Network for Computational Nanotechnology,
Purdue University is acknowledged for improving the manuscript.
\end{acknowledgments}


\end{document}